# Simulating foot-and-mouth dynamics and control in Bolivia

Nicolas C. Cardenas[1], Diego Viali dos Santos[2], Daniel Magalhães Lima[2], Hernán Oliver Daza Gutierrez[3], Daniel Rodney Gareca Vaca[3], Gustavo Machado[1]

[1]Department of Population Health and Pathobiology, College of Veterinary Medicine, North Carolina State University, Raleigh, NC, USA.

[2]Pan American Center for Foot-and-Mouth Disease and Veterinary Public Health, Pan American Health Organization, Rio de Janeiro, Brazil.

[3]Servicio Nacional de Sanidad Agropecuaria e Inocuidad Alimentaria (SENASAG). Santa Cruz de la Sierra, Bolivia.

Corresponding author: gmachad@ncsu.edu

## Abstract

Examining the dissemination dynamics of foot-and-mouth disease (FMD) is critical for revising national response plans. We developed a stochastic SEIR metapopulation model to simulate FMD outbreaks in Bolivia and explore how the national response plan impacts the dissemination among all susceptible species. We explored variations in the control strategies, mapped high-risk areas, and estimated the number of vaccinated animals during the reactive ring vaccination. Initial outbreaks ranged from 1 to 357 infected farms, with control measures implemented for up to 100 days, including control zones, a 30-day movement ban, depopulation, and ring vaccination. Combining vaccination (50-90 farms/day) and depopulation (1-2 farms/day) controlled 60.3% of outbreaks, while similar vaccination but higher depopulation rates (3-5 farms/day) controlled 62.9% and eliminated outbreaks nine days faster. Utilizing depopulation alone controlled 56.76% of outbreaks, but had a significantly longer median duration of 63 days.

Combining vaccination (25-45 farms/day) and depopulation (6-7 farms/day) was the most effective, eliminating all outbreaks within a median of three days (maximum 79 days). Vaccination alone controlled only 0.6% of outbreaks and had a median duration of 98 days. Ultimately, results showed that the most effective strategy involved ring vaccination combined with depopulation, requiring a median of 925,338 animals to be vaccinated. Outbreaks were most frequent in high-density farming areas such as Potosí, Cochabamba, and La Paz. Our results suggest that emergency ring vaccination alone can not eliminate FMD if reintroduced in Bolivia, and combining depopulation with vaccination significantly shortens outbreak duration. These findings provide valuable insights to inform Bolivia's national FMD response plan, including vaccine requirements and the role of depopulation in controlling outbreaks.

## Introduction

Foot-and-mouth disease (FMD) is a contagious viral disease that impacts cloven-hoofed animals such as cattle, swine, small ruminants, and wildlife (Paton et al., 2018). In the U.K. and the Netherlands during the 2001 FMD outbreaks, more than 6.7 million animals were slaughtered, including healthy sheep, goats, and pigs, as a precautionary measure (Bouma et al., 2003; Kitching, 2005), generating substantial economic consequences given animal loss, decreasing animal milk and beef production and affecting other industries such as tourism, with estimated costs ranging from 2.7 to 3.2 billion (Chanchaidechachai et al., 2021).

In recent years, South American countries have significantly advanced FMD control and eradication, achieving notable results (PANAFTOSA-OPS/OMS, 2021; Organización Panamericana de la Salud, 2024). For instance, Bolivia has made remarkable progress in controlling FMD. As of 2024, Bolivia, including the Department of Beni and the northern part of the Department of La Paz, has been recognized as the latest FMD-free zone. Bolivia stopped preventative vaccination in 2023 to become an FMD-free country. In contrast, the last outbreaks occurred in Colombia in 2017 and 2018 (Gomez et al., 2019). Since then, no outbreaks have been officially reported in Latin America, despite Venezuela's not achieving FMD-free status by WOAH (PANAFTOSA-OPS/OMS, 2021). However, the reintroduction of FMD remains a significant threat given the susceptibility of the cloven animal population to FMD in Latin America, as evidenced by the massive 2001 outbreak that affected 2,027 farms in Uruguay (Iriarte et al., 2023).

Mathematical models have been used to simulate FMD outbreaks, allowing for effective testing control and elimination plans' effectiveness (Pomeroy et al., 2017; Kirkeby et al., 2021). Simulation models have also been used to identify outbreak spatial dissemination, and estimate the number of vaccine doses; and estimate the depopulation capacity (Rushton, 2008; Australian Bureau of Agricultural and Resource Economics and Sciences (ABARES), 2019). Mechanistic mathematical modeling provides valuable insights into how the disease may unfold under various conditions, allowing animal health authorities to plan accordingly for outbreak potential and estimate necessary resources (Tildesley and Keeling, 2008; Cespedes Cardenas et al., 2024; Pesciaroli et al., 2025).

We begin with analyzing the Bolivian national livestock farms' spatial distribution and animal movement data analysis to determine species densities by regions used in selecting farms for seed infection. We then extended our multi-host, single-pathogen, multiscale model MHASpread model presented by (Cespedes Cardenas et al., 2024), to i) simulate the spread of FMD in Bolivia and evaluate the effectiveness of various control strategies and ii) map the spatial distribution patterns of outbreaks, and iii) based on the results, estimate the number of vaccinated animals under each scenario conditions simulated.

## Materials and methods

### *Population data*

We used data from the Bolivian livestock farms from January 1, 2023, to July 25, 2024, provided by the Servicio Nacional de Sanidad Agropecuaria e Inocuidad Alimentaria (SENASAG) in Bolivia (SENASAG). The dataset included information on 218,143 farms, 183,752 cattle, 309 buffalo, 70,083 swine, and 123,011 sheep and goats. We grouped cattle and buffalo farms into one group named hereafter as "bovines", while sheep and goats farms were categorized as "small ruminants". After excluding 4,034 farms (1.8%) due to missing geographical coordinates or number of heads, the final dataset consisted of 214,109 farms (Figure 1), 169,426 bovine farms, 53,313 swine farms, and 121,112 small ruminant farms.

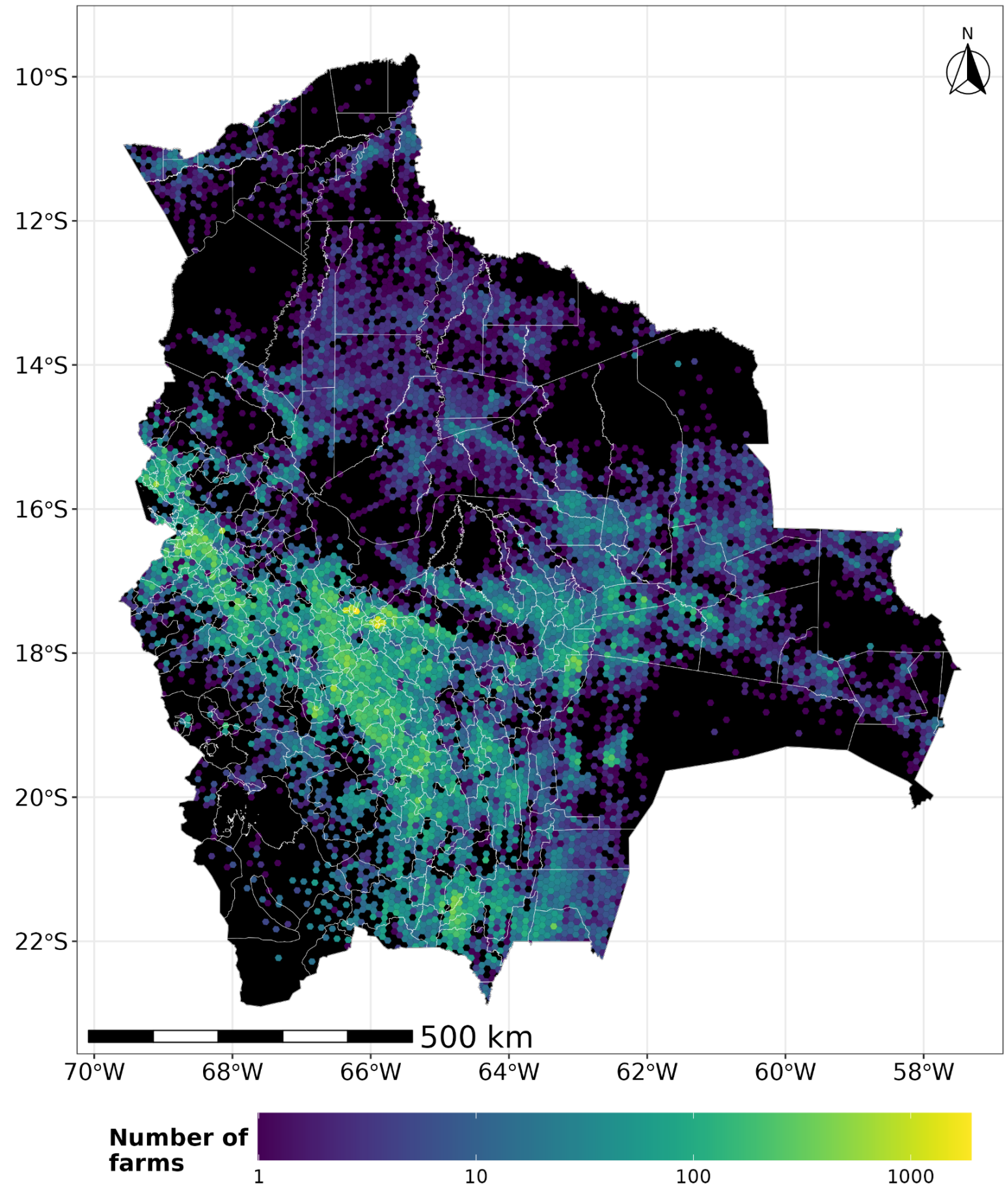

**Figure 1. Bolivia's livestock density distribution**. 10 $km^2$ hexagon polygon with total livestock population overlayed with municipalities (white borders).

*Animal movement data*

The SENASAG dataset recorded 374,268 unique animal movements from January 1, 2023, to July 25, 2024, including farm-to-farm and farm-to-slaughterhouse movements. After excluding 28,181 records (7.5%) due to missing information, including a) zero animals moved, b) identical origin and destination, c) movements from or to farms not present in the population database or out-of-state premises, our model analyzed 346,087 valid farm-to-farm and farm-to-slaughterhouse movements, corresponding to 7,806,963 animals. The distribution of movements by species, categorized as farm-to-farm and farm-to-slaughterhouse, is shown in Table 1.

**Table 1.** Number of farm-to-farm and farm-to-slaughterhouse movements and animals count by host species in Bolivia from January 1, 2023, to July 25, 2024

| Species | Movement type | Number of movements | % Movements | Number of animals | % Animals |
|---|---|---|---|---|---|
| **Bovine** | Farm-to-farm | 156,708 | 95.85% | 3,738,579 | 83.66% |
| **Small ruminants** | Farm-to-farm | 3,219 | 1.97% | 146,726 | 3.28% |
| **Swine** | Farm-to-farm | 3,557 | 2.18% | 582,756 | 13.05% |
| **All species** | Farm-to-farm | 163,484 | 100.00% | 4,468,061 | 100.00% |
| **Bovine** | Farm-to-slaughterhouse | 150,053 | 82.19% | 1,759,277 | 52.72% |
| **Small ruminants** | Farm-to-slaughterhouse | 65 | 0.04% | 408 | 0.01% |
| **Swine** | Farm-to-slaughterhouse | 32,485 | 17.79% | 1,579,217 | 47.27% |
| **All species** | Farm-to-slaughterhouse | 182,603 | 100.00% | 3,338,902 | 100.00% |

*Model description*

We extended our multi-host, single-pathogen model to simulate FMD outbreaks and control measures and implemented the MHASpread (version 3.0) (https://github.com/machado-lab/MHASPREAD-model) described elsewhere (Cespedes Cardenas and Machado, 2024; Cespedes Cardenas et al., 2024). The supplementary material contains a detailed description of the model methodology and parametrization. Briefly, MHASpread consists of a Suceptibel-Exposed-Infected-Recovered (SEIR) compartment model in which the population, including bovine, swine, and small ruminants, is divided into compartments. The disease transmission is modeled using species-specific transmission probabilities, reflecting the unique transmission dynamics between species (Supplementary Material Table S1) and disease progression across compartments for each species within each farm population (Supplementary Material Table S2). The model assumes a homogeneous mixing of species within farms while considering the movement data for between-farm dynamics of animal movements. Our model accounts for movements from farms to slaughterhouses, where transported animals are permanently removed from the simulation. The between-farm spatial transmission process was modeled as a transmission kernel, defining the probability of disease spread based on the distance between infected and non-infected farms. The spatial transmission kernel likelihood of disease transmission decreases as the between-farm distance increases to a maximum of 40 km (Boender et al., 2010; Boender and Hagenaars, 2023). The transmission kernel simplifies the complex dynamics of between-farm disease spread by theoretically covering all forms of transmission within this distance limit (Ferguson et al., 2001).

*FMD spread and control actions*

*Initial conditions*

A representative sample was drawn from the Bolivian livestock population of 214,109 farms to serve as the initial infected farms. This sample was selected using a multistage, stratified approach, stratifying by species among the nine Bolivian departments. The sample size was calculated with an assumed prevalence of 50% to maximize the total size with a 95% confidence level and a 1.1% margin of error, which results in 1,061 sampled farms (geolocation distribution in Supplementary Material Figure S1). These selected farms initiated the infection with five infected animals per farm, regardless of the farm's population size. The species targeted for seeding infection was bovine; however, small ruminants were initially infected if no bovines were present on a given farm. Five swine were initially infected if neither bovines nor small ruminants were present. The initial outbreak model was run for 20 days (Figure 2), which generated outbreaks of 1 to 357 infected farms.

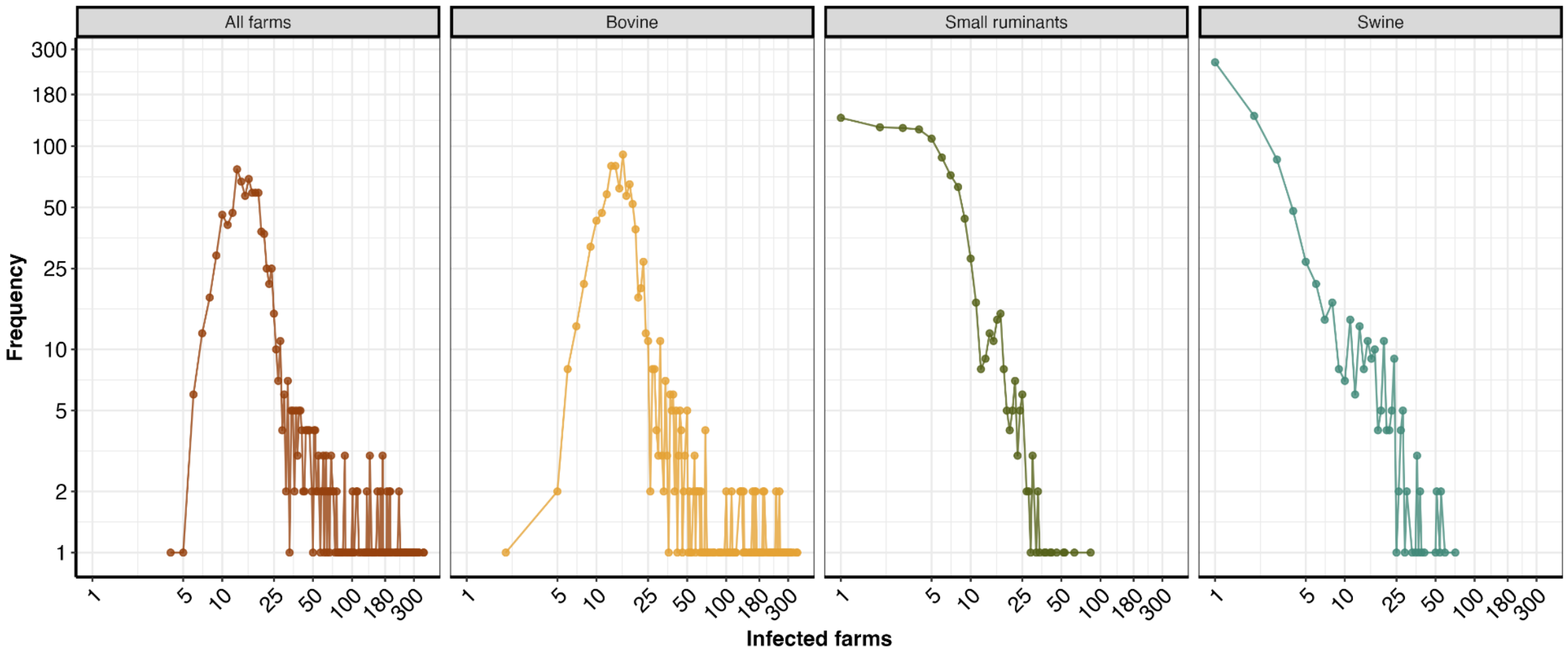

**Figure 2. Initial outbreak size.** Distribution of the frequencies of an initial number of infected farms by host species in 20 days from initial infection.

*Control strategies scenarios*

We outline five elimination scenarios based on strategies outlined in the national response plan by SENASAG (SENASAG, 2015). These strategies were developed in collaboration with SENASAG animal officials and selected according to local capacities and the organizational structure of the official veterinary service. Briefly, these actions include the depopulation of infected farms, emergency vaccination of infected and uninfected farms, animal movement standstill, trace-back, and the establishment of three control zones around infected farms at the following distances: a 3 km infected zone, a 7 km buffer zone, and 15 km surveillance zone (Supplementary Material Figure S2, Table 1). Depopulation: this measure is carried out within infected zones in which infected farms are depopulated, we impose a maximum number of farms that can be eliminated on a daily basis (Table 1), with larger farms being depopulated first. Once a farm was depopulated, its entire animal population was removed, and the farm(s) were excluded from the simulations. Once depopulation's daily capacity is reached, not depopulated farms are scheduled to be depopulated the following day (Table 1). Emergency vaccination was administered in both the infected and buffer zones, which started 15 days after the initial detection and start of control actions to account for the preparation time for the vaccines. Depending on the scenario, infected farms can be vaccinated or not. Emergency vaccinations were administered in both the infected and buffer zones depending on the scenario configuration (Table 1). Once farms in the infected zone are vaccinated, any remaining vaccine doses are used to immunize farms in the buffer zone. If a farm can not be vaccinated within a day due to capacity constraints, it is scheduled for vaccination the following day(s) (Table 1). In each simulation, we randomly chose a vaccine efficacy level. This level determined the percentage of vaccinated animals that became immune to FMD within 15 days, with the efficacy levels being 80%, 90%, or 100%. (Backer et al., 2012; Ulziibat et al., 2023a). Refer to the "Vaccination" section in the Supplementary Material for more detailed information on the vaccination methodology.

Control measures were initiated on day 20 after the outbreak onset, for the initial day of control actions it was assumed that a 10% proportion of farms would be detected. For example, if 100 farms were infected, 10 were detected, and when fewer than one farm was infected, the model rounded to one detected farm. For the following days of control actions, the detection rate varied based on the number of farms within the control zone(s) and the total number of infected farms. When fewer farms remain under surveillance, the proportion of infected farms within this population increases, making detection more likely (Supplementary Figure S3). In contrast, when more farms are under surveillance, but the infection is more dispersed, detection may take longer due to lower disease prevalence across the monitored farms (Supplementary Material and Supplementary Material Figure S3).

**Table 2.** Description of elimination scenarios and parameters.

| Parameter | Scenario 1 | Scenario 2 | Scenario 3 | Scenario 4 | Scenario 5 |
|---|---|---|---|---|---|
| **Initial Conditions** | | | | | |
| Infected farm | random | random | random | random | random |
| The initial simulation day | random | random | random | random | random |
| **Detection** | | | | | |
| Days of control action | 120 | 120 | 120 | 120 | 120 |
| **Control zone(s) radii in kilometers** | | | | | |
| Infected zone | 3 | 3 | 3 | 3 | 3 |
| Buffer zone | 5 | 5 | 5 | 5 | 5 |
| Surveillance zone | 7 | 7 | 7 | 7 | 7 |
| **Movement restriction** | | | | | |
| Standstill (days) | 30 | 30 | 30 | 30 | 30 |
| The standstill in the infected zone | F | F | F | T | F |
| The standstill in the buffer zone | F | F | F | T | F |
| The standstill in the surveillance zone | T | T | T | T | T |
| Direct contact with infected/detected farms | F | F | F | F | F |
| Traceback duration | 1 | 1 | 1 | 2 | 1 |
| **Depopulation** | | | | | |
| Limit of farms depopulated per day | 1-2 | 1-2 | 3-5 | 0 | 6-7 |
| Depopulation in the infected zone | F | F | F | F | F |
| Only depopulate infected farms | T | T | T | T | T |
| **Vaccination** | | | | | |
| Days to achieve immunity | 15 | 15 | NA | 15 | 15 |
| Number of farms vaccinated in the buffer zone per day | 25-45 | 45-60 | NA | 25-45 | 25-45 |
| Number of farms vaccinated in the infected per day | 25-45 | 45-60 | NA | 25-45 | 25-45 |
| Proportion of vaccine efficacy | 0.8, 0.9, 1 | 0.8, 0.9, 1 | NA | 0.8, 0.9, 1 | 0.8, 0.9, 1 |
| Vaccination of swine | F | F | NA | NA | F |
| Vaccination of bovines | T | T | NA | T | T |
| Vaccination of small ruminants | F | F | NA | NA | F |
| Vaccination in the infected zone | T | T | NA | NA | T |
| Vaccination in the buffer zone | F | T | NA | NA | F |
| Vaccination delay (days) | 15 | 15 | NA | NA | 15 |
| Vaccination in infectious farms | F | T | NA | T | T |

"T" denotes True, indicating that the control action parameter is applied, and "F" is false when control actions were not applied. "NA" signifies Not Applicable, meaning the parameter configuration is irrelevant to the current scenario.

We further classified outbreaks into complete and incomplete outbreak elimination, named "controlled" if all infected farms were eliminated within 100 days of the start of control actions.

Simulation(s) was deemed “not controlled” if either of the following occurred: 1) the number of infected farms surpassed 400 at any point, or 2) control measures remained active for over 100 days while more than one farm was still infected.

*Spatial spread analysis and mapping*

Using the simulation results, we quantified how often each farm was infected across all simulations, regardless of the control scenarios. We then grouped these results by municipality by summing the infections for each farm and calculating the percentage of infected farms relative to the total number of farms in each municipality.

*Number of vaccinated animals*

We calculated the daily number of vaccinated animals; vaccination is used in infected and buffer zones, but not in surveillance zones. We assumed the vaccine produced protection for six months with a single dose (Ulziibat et al., 2023b) and vaccinated animals remained alive and non-susceptible until the end of their production cycle under the vaccinate-to-live strategy scenarios one, two, four, and five (Table 1). In our model, depopulation is scheduled before vaccination, meaning that farms that have already been depopulated are no longer considered for immunization. However, infected farms are eligible for vaccination in scenarios 2, 4, and 5. In these cases, animals are counted as vaccinated, even if the farm is depopulated the following day (Table 1).

*Software*

The language software used to develop the MHASpread model (Cespedes Cardenas and Machado, 2024) and create graphics, tables, and maps was R v. 4.2.3 (R Core Team, 2024) and Python v. 3.8.12, R utilizing the following packages: sampler (Baldassaro, 2019), tidyverse (Wickham et al., 2019), sf (Pebesma, 2018), doParallel (Corporation and Weston, 2022), lubridate (Grolemund and Wickham, 2011) and Python v. 3.8.12 with the following packages: Numpy (Harris et al., 2020), Pandas (McKinney, 2010), and SciPy (Virtanen et al., 2020).

## Results

### *Distribution of the number of infected farms over time*

Figure 3 represents the outbreak distributions' results over time, distinguishing between "controlled" and "not controlled" outbreaks. Here, scenario 5 eliminated all outbreaks with a median outbreak duration of three days, with a maximum of 79 days. In contrast, scenario 4, with a lower vaccination coverage and no depopulation eliminated 0.6% of the outbreaks and the median outbreak duration of 98 days. Both scenario 1 and scenario 2 demonstrated similar effectiveness, controlling 60.3% to 62.9% of outbreaks, respectively. However, in scenario 2, the median time to eliminate outbreaks was nine days less than in scenario 1. Scenario 3, the scenario in which vaccination was not used and 3-5 farms were depopulated daily was effective, controlling 56.76% of outbreaks; nevertheless, it had a significantly longer median outbreak duration of 63 days (Figure 3 and Table 2).

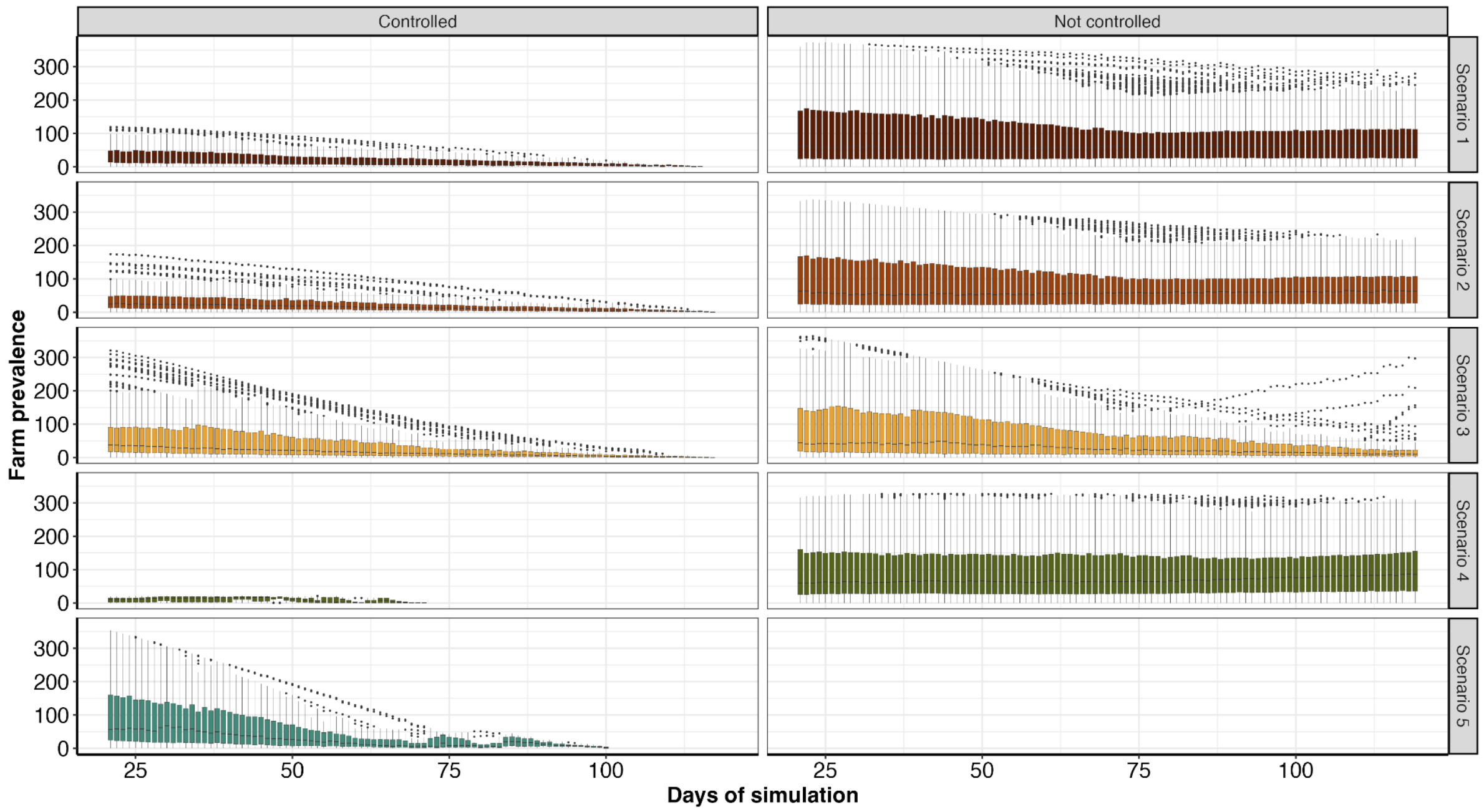

**Figure 3. Box plot of epidemic trajectories by control scenarios**. Each card represents the distribution of outbreaks that were controlled and those that were not controlled by the different control scenarios. The y-axis denotes the number of infected farms, while the x-axis represents the simulation days.

*Spatial spread analysis and mapping*

Figure 4 illustrates the percentage of infected farms by municipality. The municipalities with the highest infection rates were Tinguipaya (3.69%), Tapacarí (2.78%), and Méndez (2.57%). While, Detosí, Cochabamba, and La Paz department regions accounted for 25.3%, 23.5%, and 18.6% of these infections, respectively. Santa Cruz and Tarija also experienced notable percentages, with 17.4% and 10.9%, followed by Chuquisaca (5.45%), Pando (4.77%), Oruro (4.33%), and Beni (4.02%).

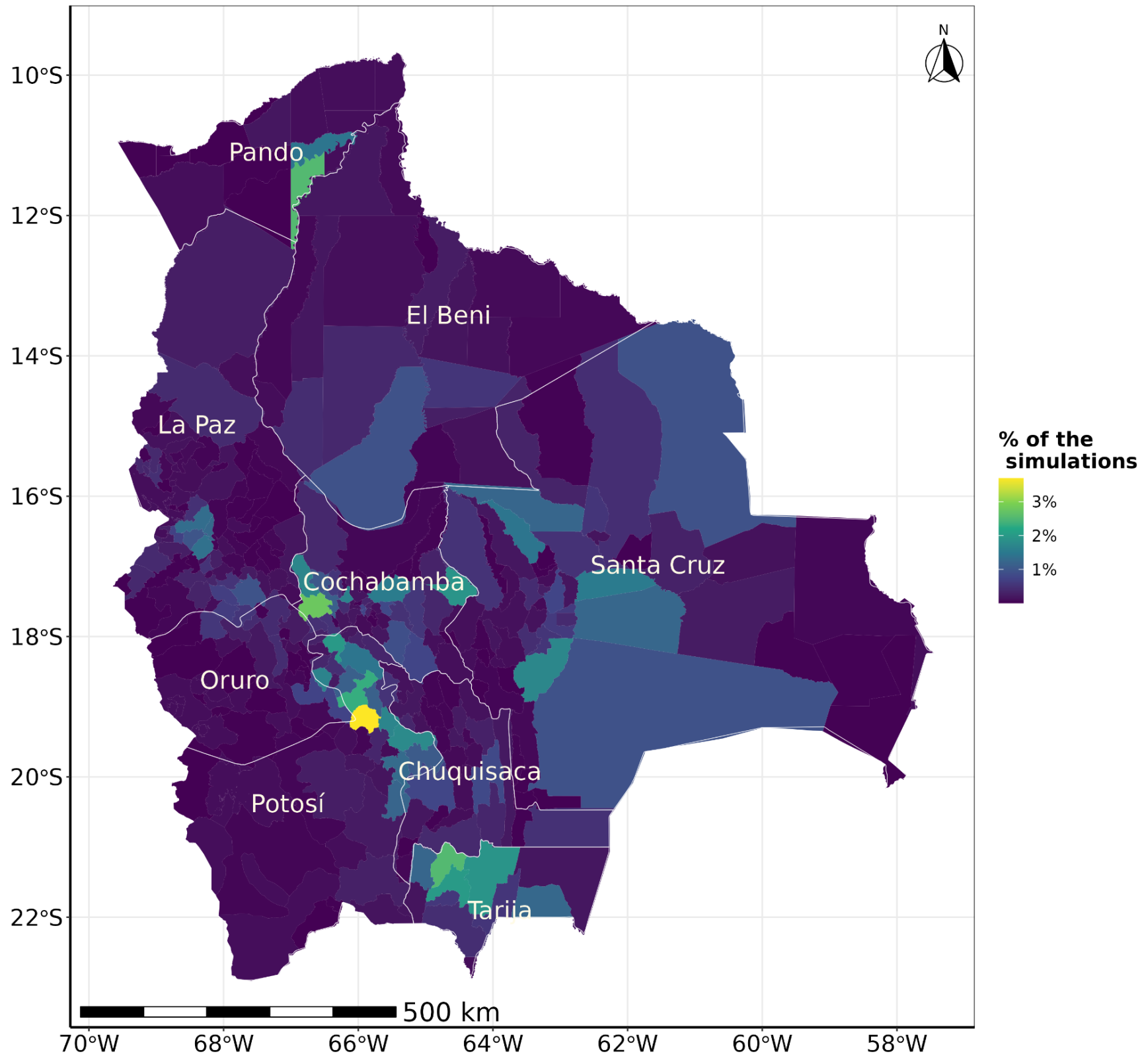

**Figure 4. Percentage of infected farms by the municipality.** The color represents the percentage of infected farms over all simulations at the municipality level. The gray lines represent the department of Bolivia.

*Number of vaccinated animals*

Figure 5 presents the daily distribution of vaccinated animals over time. The median number of daily vaccinated animals was 1,962 (IQR: 726 to 6,752), with daily values ranging from 0 to 492,492. Scenario 5 exhibited the highest daily vaccinated animals with a median of 42,705 (IQR: 22,414 to 97,500), while scenario 1 showed the lowest median value at 2,564 (IQR: 1,020 to 9,608). Both scenarios 2 and 4 yielded intermediate results, with a daily median of 2,832 (IQR: 1,040 to 10,380) and 1,424 (IQR: 528 to 4,320), respectively (Table 3 and Figure 5).

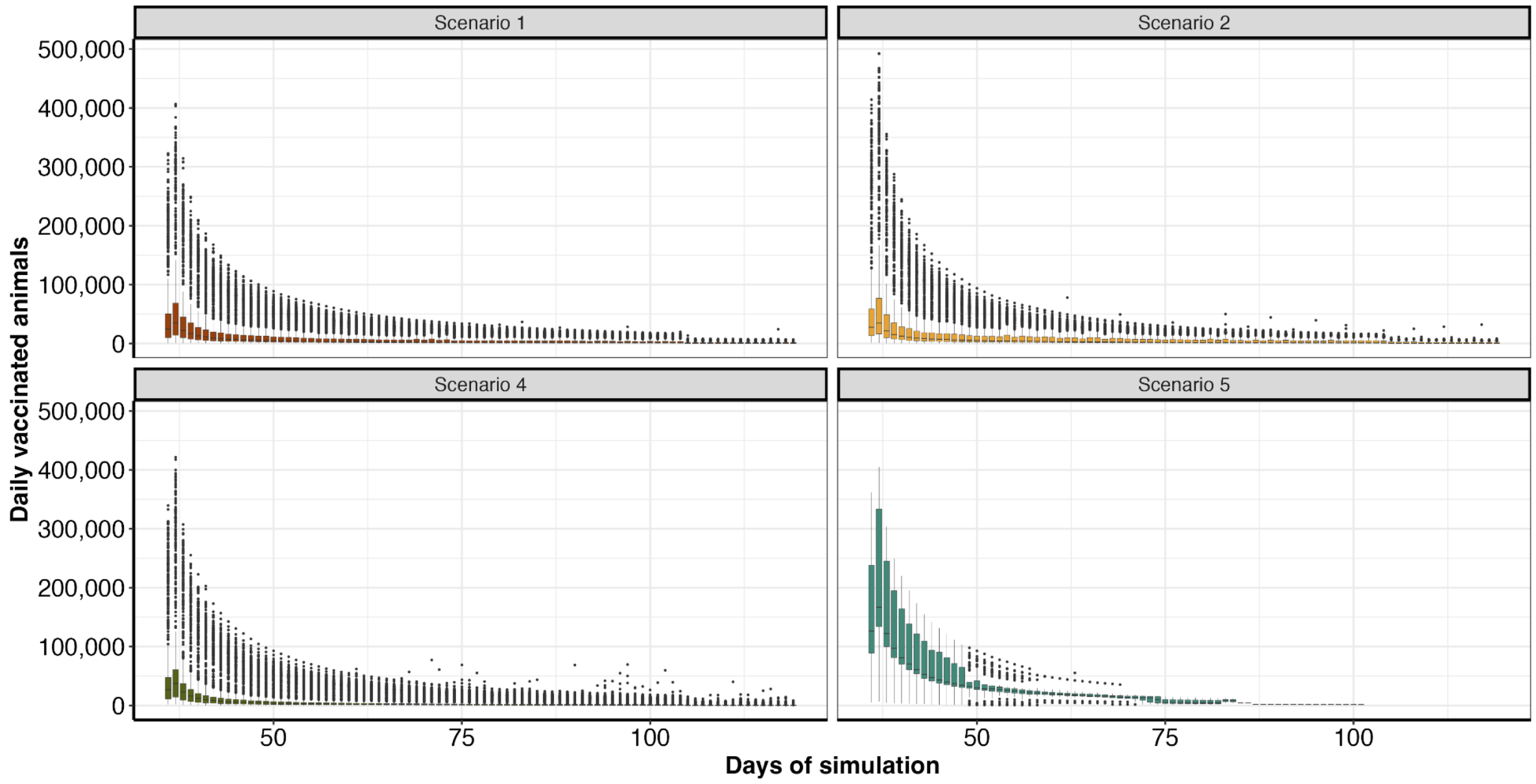

**Figure 5. Distribution of the daily number of vaccinated animals by scenario over time**

**Table 3.** Performance metrics of simulation scenarios for disease control. The metrics include the daily vaccination rate, depopulation rate, percentage of controlled outbreaks, the median duration of control actions (including interquartile range and maximum values), and the total number of vaccine animals administered.

| Scenario | Vaccination rate/day (infected +buffer zone) | Depopulation rate/day | Percentage of controlled outbreaks % | Control actions duration (median, IQR, max) | Total number of vaccinated animals (median, IQR, max) |
|---|---|---|---|---|---|
| 1 | 50 to 90 | 1 to 2 | 60.3% | 37 (IQR: 11 - 98, max: 98) | 236,526 (IQR: 96,777 - 468,004, max: 4,676,797) |
| 2 | 90 to 120 | 1 to 2 | 62.9% | 28 (IQR: 11 - 98, max: 98) | 236,389 (IQR: 99,081 - 486,604, max: 5,016,391) |
| 3 | 0 | 3 to 5 | 56.8% | 63 (IQR: 29 - 97, max: 98) | - |
| 4 | 50 to 90 | 0 | 0.6% | 98 (IQR: 98 - 98, max: 98) | 260,305 (IQR: 140,312 - 445,072, max: 4,811,808) |
| 5 | 50 to 90 | 6 to 7 | 100% | 3 (IQR: 2 - 6, max: 79) | 925,338 (IQR: 404,322 - 1,889,406, max: 3,691,642) |

**Discussion**

In the present study, we used our MHASpread model framework (Cespedes Cardenas et al., 2024; Cardenas et al., 2025) to simulate FMD outbreaks in Bolivia to estimate the outbreak size, map the spatial distribution of the outbreaks and estimate the number of vaccinated animals. We demonstrated that combining vaccination and depopulation is the most effective strategy for controlling outbreaks, eliminating up to 100% of outbreaks within a median duration of three days. In contrast, single-control approaches, depopulation alone controlled only 56.76% of outbreaks with a median duration of 63 days. Low vaccination coverage without depopulation controlled just 0.6% of outbreaks, with the longest median duration of 98 days. Depopulation was the most effective measure in reducing outbreaks, while emergency vaccination reduced the time spent on control actions by vaccinating in a median between 236,526 and 925,338 bovines in 100 days of simulated control actions.

Regions with a high livestock density were correlated with our simulated epidemic hotspots (Figure 4). Understanding the impact of initially infected farms and quantifying the number of affected farms could be used to assess how changes in production patterns may influence future disease dissemination risks. Our results indicate that the regions of Detosí, Cochabamba, and La Paz accounted for 25.3%, 23.5%, and 18.6% of these infections, respectively. This suggests that host density may play an important, particularly significant role in mediating the severity of epidemics in Bolivia (Meadows et al., 2018; Seibel et al., 2024). Thus, our results indicated that for regions with higher farm densities, such as the central part of Bolivia, Potosí, Cochabamba, and La Paz (Figure 1), would be ideal locations for stockpiling the necessary resources to respond to a future FMD introduction. These findings align with previous studies showing a correlation between higher farm density and increased infection rates (Hayama et al., 2013; Halasa et al., 2020; Marschik et al., 2021; Seibel et al., 2024; Pesciaroli et al., 2025).

We demonstrated that vaccination and depopulation of infected farms were the most impactful control actions and are part of Bolivia's contingency plans (SENASAG, 2015). Vaccinating between 50 and 90 farms and depopulating from 6 to 7 farms daily eliminated 100% of the outbreaks within three days of the initial detection. However, this scenario required a median of 255% more vaccinated animals, which is about more than three times as many as any other simulated scenario. In contrast, a scenario with only vaccination eliminates 1% of the outbreaks, highlighting that vaccination alone is insufficient to control most outbreaks under simulated conditions. Moreover, in scenarios where the depopulation rate remained constant for the 100 days of simulated control action (1 to 2 farms per day) and included vaccination, our results indicated that increasing the vaccination rate from 50-90 to 90-120 farms daily reduced the median duration of the outbreak by nine days demonstrating the benefits of emergency vaccination when combined with depopulation of infected farms. This is because vaccinated animals are less likely to become infectious, reflecting an immune response that reduces virus shedding (Stenfeldt et al., 2016) and provides some level of protection to unvaccinated animals, as the spread of the disease is significantly reduced (Orsel and Bouma, 2009; Singh et al., 2019). Other studies found similar patterns, indicating that vaccination may be effective in reducing the duration and size of FMD outbreaks (Tomassen et al., 2002; Keeling et al., 2003; Dürr et al., 2014; Roche et al., 2014; Garner et al., 2016). Indeed, our findings are aligned with studies from Australia, Canada, New Zealand, and the United States, where emergency vaccination was linked to a reduction in the number of infected farms and a shorter outbreak duration (Traulsen et al., 2011; Porphyre et al., 2013; Roche et al., 2015; Sanson et al., 2017; Rawdon et al., 2018).

We remark that the limitation of simulating FMD dynamics without historical outbreaks limits our ability to perform model calibration and rely on taking parameters from the literature (Cespedes Cardenas et al., 2024). Furthermore, future work will estimate the cost of all these interventions, which the authors consider essential for designing decision-making tools.

## Conclusion

Our model indicated that vaccination and depopulation would be required to eliminate FMD if reintroduced in Bolivia and the total number of vaccine doses needed could reach 4,811,808 along with depopulation. Increasing the number of vaccinated farms from 90 to 120 while maintaining depopulation at two farms per day improved the percentage of controlled outbreaks by 2%, from 60.3% to 62.9%. In contrast, increasing the depopulation rate to five farms per day without vaccination resulted in 56.8% of controlled outbreaks, only 4% less effective than combined strategies involving vaccination and depopulation. These findings highlight that while vaccination contributes to outbreak control, its effectiveness is limited without concurrent depopulation. Ultimately, the likelihood of FMD eradication depends on sufficient vaccine doses, timely deployment, and the capacity to depopulate one to two farms daily, regardless of population size. Early detection and rapid implementation of control measures remain critical for successful containment.

**Acknowledgments**

We would like to thank the animal health officials of Servicio Nacional de Sanidad Agropecuaria e Inocuidad Alimentaaria (SENASAG): Liliana Justiniano Ruiz, Olquer Hugo Calle Guzmán, Wilfredo Wils Alvarez Garzon, Ronald Orellana Orellana, Eduardo Tapia Maigua Jose Luis Cruz Garcia, and Ronny Salvatierra for their constant contributions to discussions regarding the state FMD response plan in Bolivia.

**Authors' contributions**

NCC and GM conceived the study. NCC, DM, DVDS, and GM participated in the design of the study. HODG, DM, and DRGV coordinated the data collection. NCC and DM conducted data processing and cleaning, designed the model, and simulated scenarios. NCC conducted the formal coding and initial draft. GM, DVDS, and NCC wrote and edited the manuscript. All authors discussed the results and critically reviewed the manuscript. GM secured the funding.

**Conflict of interest**

All authors confirm that there are no conflicts of interest to declare.

**Ethical statement**

The authors confirm the ethical policies of the journal, as noted on the journal's author guidelines page. Since this work did not involve animal sampling nor questionnaire data collection by the researchers, there was no need for ethics permits.

**Data Availability Statement**

The data that support the findings of this study are not publicly available and are protected by confidential agreements, therefore, are not available.

**Funding**

This work was supported by the Machado Laboratory at Department of Population Health and Pathobiology, College of Veterinary Medicine, North Carolina State University, Raleigh, NC, USA.

**Supporting Information**

Filename: Supplementary Material Bolivia FMD.pdf document, 1.4MB

Description:

Supporting Information Material and Methods model formulation and description

Supporting Information Table S1. The distribution of each host-to-host transmission coefficient (β) per animal$_{-1}$ day$_{-1}$.

Supporting Information Table S2. The within-farm distribution of latent and infectious FMD parameters for each species.

Supporting Information Figure S1. The grid density of the sampled farms to seed initial outbreaks.

Supporting Information Figure S2. Control zones mapping.

Supporting Information Figure S3. Percentage of detected farms according to prevalence and population size.

**References**

Australian Bureau of Agricultural and Resource Economics and Sciences (ABARES), 2019:

Potential socio-economic impacts of an outbreak of foot-and-mouth disease in Australia. .

Backer, J.A., T.J. Hagenaars, G. Nodelijk, and H.J.W. van Roermund, 2012: Vaccination against foot-and-mouth disease I: Epidemiological consequences. *Prev. Vet. Med.* **107**, 27–40, DOI: 10.1016/j.prevetmed.2012.05.012.

Baldassaro, M., 2019: sampler: Sample Design, Drawing & Data Analysis Using Data Frames. . CRAN.

Boender, G.J., and T.J. Hagenaars, 2023: Common features in spatial livestock disease transmission parameters. *Sci. Rep.* **13**, 3550, DOI: 10.1038/s41598-023-30230-w.

Boender, G.J., H.J.W. van Roermund, M.C.M. de Jong, and T.J. Hagenaars, 2010: Transmission risks and control of foot-and-mouth disease in The Netherlands: Spatial patterns. *Epidemics* **2**, 36–47, DOI: 10.1016/j.epidem.2010.03.001.

Bouma, A., A.R.W. Elbers, A. Dekker, A. de Koeijer, C. Bartels, P. Vellema, P. van der Wal, E.M.A. van Rooij, F.H. Pluimers, and M.C.M. de Jong, 2003: The foot-and-mouth disease epidemic in The Netherlands in 2001. *Prev. Vet. Med.* **57**, 155–166, DOI: 10.1016/S0167-5877(02)00217-9.

Cardenas, N.C., T.C. de Menezes, A.M. Countryman, F.P.N. Lopes, F.H.S. Groff, G.M. Rigon, M. Gocks, and G. Machado, 2025: Integrating epidemiological and economic models to estimate the cost of simulated foot-and-mouth disease outbreaks in Brazil. DOI: 10.48550/ARXIV.2502.07083. arXiv.

Cespedes Cardenas, N., F. Amadori Machado, C. Trois, V. Maran, A. Machado, and G. Machado, 2024: Modeling foot-and-mouth disease dissemination in Rio Grande do Sul, Brazil and evaluating the effectiveness of control measures. *Front. Vet. Sci.* **11**.

Cespedes Cardenas, N., and G. Machado, 2024: MHASPREAD model [Online] Available at

https://github.com/machado-lab/MHASPREAD-model (accessed September 16, 2024).

Chanchaidechachai, T., M.C.M. de Jong, and E.A.J. Fischer, 2021: Spatial model of foot-and-mouth disease outbreak in an endemic area of Thailand. *Prev. Vet. Med.* **195**, 105468, DOI: 10.1016/j.prevetmed.2021.105468.

Corporation, M., and S. Weston, 2022: doSNOW: Foreach Parallel Adaptor for the "snow" Package. . CRAN.

Dürr, S., C. Fasel-Clemenz, B. Thür, H. Schwermer, M.G. Doherr, H.Z. Dohna, T.E. Carpenter, L. Perler, and D.C. Hadorn, 2014: Evaluation of the benefit of emergency vaccination in a foot-and-mouth disease free country with low livestock density. *Prev. Vet. Med.* **113**, 34–46, DOI: 10.1016/j.prevetmed.2013.10.015.

Ferguson, N.M., C.A. Donnelly, and R.M. Anderson, 2001: Transmission intensity and impact of control policies on the foot and mouth epidemic in Great Britain. *Nature* **413**, 542–548, DOI: 10.1038/35097116.

Garner, M.G., N. Bombarderi, M. Cozens, M.L. Conway, T. Wright, R. Paskin, and I.J. East, 2016: Estimating Resource Requirements to Staff a Response to a Medium to Large Outbreak of Foot and Mouth Disease in Australia. *Transbound. Emerg. Dis.* **63**, e109–e121, DOI: 10.1111/tbed.12239.

Grolemund, G., and H. Wickham, 2011: Dates and Times Made Easy with {lubridate}. *J. Stat. Softw.* **40**, 1--25.

Halasa, T., M.P. Ward, and A. Boklund, 2020: The impact of changing farm structure on foot-and-mouth disease spread and control: A simulation study. *Transbound. Emerg. Dis.* **67**, 1633–1644, DOI: 10.1111/tbed.13500.

Harris, C.R., K.J. Millman, S.J. van der Walt, R. Gommers, P. Virtanen, D. Cournapeau, E.

Wieser, J. Taylor, S. Berg, N.J. Smith, R. Kern, M. Picus, S. Hoyer, M.H. van Kerkwijk, M. Brett, A. Haldane, J.F. del Río, M. Wiebe, P. Peterson, P. Gérard-Marchant, K. Sheppard, T. Reddy, W. Weckesser, H. Abbasi, C. Gohlke, and T.E. Oliphant, 2020: Array programming with NumPy. *Nature* **585**, 357–362, DOI: 10.1038/s41586-020-2649-2.

Hayama, Y., T. Yamamoto, S. Kobayashi, N. Muroga, and T. Tsutsui, 2013: Mathematical model of the 2010 foot-and-mouth disease epidemic in Japan and evaluation of control measures. *Prev. Vet. Med.* **112**, 183–193, DOI: 10.1016/j.prevetmed.2013.08.010.

Iriarte, M.V., J.L. Gonzáles, E. De Freitas Costa, A.D. Gil, and M.C.M. De Jong, 2023: Main factors associated with foot-and-mouth disease virus infection during the 2001 FMD epidemic in Uruguay. *Front. Vet. Sci.* **10**, 1070188, DOI: 10.3389/fvets.2023.1070188.

Keeling, M.J., M.E.J. Woolhouse, R.M. May, G. Davies, and B.T. Grenfell, 2003: Modelling vaccination strategies against foot-and-mouth disease. *Nature* **421**, 136–142, DOI: 10.1038/nature01343.

Kirkeby, C., V.J. Brookes, M.P. Ward, S. Dürr, and T. Halasa, 2021: A Practical Introduction to Mechanistic Modeling of Disease Transmission in Veterinary Science. *Front. Vet. Sci.* **7**, 546651, DOI: 10.3389/fvets.2020.546651.

Kitching, R., 2005: Global epidemiology and prospects for control of foot-and-mouth disease. *Foot--Mouth Dis. Virus*133–148.

Marschik, T., I. Kopacka, S. Stockreiter, F. Schmoll, J. Hiesel, A. Höflechner-Pöltl, A. Käsbohrer, and B. Conrady, 2021: What Are the Human Resources Required to Control a Foot-and-Mouth Disease Outbreak in Austria? *Front. Vet. Sci.* **8**, 727209, DOI: 10.3389/fvets.2021.727209.

McKinney, W., 2010: Data Structures for Statistical Computing in Python. pp. 51–56. In: Walt, Stéfan van der, and Jarrod Millman (eds), *Proc. 9th Python Sci. Conf.*

Meadows, A.J., C.C. Mundt, M.J. Keeling, and M.J. Tildesley, 2018: Disentangling the influence of livestock vs. farm density on livestock disease epidemics. *Ecosphere* **9**, e02294, DOI: 10.1002/ecs2.2294.

Organización Panamericana de la Salud, 2024: COSALFA 50. Informe de la Secretaria Ex Officio para la 50a Reunión de la Comisión Sudamericana de Lucha Contra la Fiebre Aftosa p. 17. .

Orsel, K., and A. Bouma, 2009: The effect of foot-and-mouth disease (FMD) vaccination on virus transmission and the significance for the field. *Can. Vet. J. Rev. Veterinaire Can.* **50**, 1059–1063.

PANAFTOSA-OPS/OMS, 2021: Informe de Situación de los Programas de Erradicación de la Fiebre Aftosa en Sudamérica y Panamá, año 2021. .

Paton, D.J., S. Gubbins, and D.P. King, 2018: Understanding the transmission of foot-and-mouth disease virus at different scales. *Curr. Opin. Virol.* **28**, 85–91, DOI: 10.1016/j.coviro.2017.11.013.

Pebesma, E., 2018: Simple Features for R: Standardized Support for Spatial Vector Data. *R J.* **10**, 439, DOI: 10.32614/RJ-2018-009.

Pesciaroli, M., A. Bellato, A. Scaburri, A. Santi, A. Mannelli, and S. Bellini, 2025: Modelling the Spread of Foot and Mouth Disease in Different Livestock Settings in Italy to Assess the Cost Effectiveness of Potential Control Strategies. *Animals* **15**, 386, DOI: 10.3390/ani15030386.

Pomeroy, L.W., S. Bansal, M. Tildesley, K.I. Moreno-Torres, M. Moritz, N. Xiao, T.E.

Carpenter, and R.B. Garabed, 2017: Data-Driven Models of Foot-and-Mouth Disease Dynamics: A Review. *Transbound. Emerg. Dis.* **64**, 716–728, DOI: 10.1111/tbed.12437.

Porphyre, T., H.K. Auty, M.J. Tildesley, G.J. Gunn, and M.E.J. Woolhouse, 2013: Vaccination against Foot-And-Mouth Disease: Do Initial Conditions Affect Its Benefit? (Vittoria Colizza, Ed.)*PLoS ONE* **8**, e77616, DOI: 10.1371/journal.pone.0077616.

R Core Team, 2024: R: A Language and Environment for Statistical Computing. Vienna, Austria: R Foundation for Statistical Computing.

Rawdon, T.G., M.G. Garner, R.L. Sanson, M.A. Stevenson, C. Cook, C. Birch, S.E. Roche, K.A. Patyk, K.N. Forde-Folle, C. Dubé, T. Smylie, and Z.D. Yu, 2018: Evaluating vaccination strategies to control foot-and-mouth disease: a country comparison study. *Epidemiol. Infect.* **146**, 1138–1150, DOI: 10.1017/S0950268818001243.

Roche, S.E., M.G. Garner, R.L. Sanson, C. Cook, C. Birch, J.A. Backer, C. Dube, K.A. Patyk, M.A. Stevenson, Z.D. Yu, T.G. Rawdon, and F. Gauntlett, 2015: Evaluating vaccination strategies to control foot-and-mouth disease: a model comparison study. *Epidemiol. Infect.* **143**, 1256–1275, DOI: 10.1017/S0950268814001927.

Roche, S.E., M.G. Garner, R.M. Wicks, I.J. East, and K. De Witte, 2014: How do resources influence control measures during a simulated outbreak of foot and mouth disease in Australia? *Prev. Vet. Med.* **113**, 436–446, DOI: 10.1016/j.prevetmed.2013.12.003.

Rushton, J., 2008: Economic aspects of foot and mouth disease in Bolivia: -EN- -FR- La dimension économique de la fièvre aphteuse en Bolivie -ES- Aspectos económicos de la fiebre aftosa en Bolivia. *Rev. Sci. Tech. OIE* **27**, 759–769, DOI: 10.20506/rst.27.3.1837.

Sanson, R., T. Rawdon, K. Owen, K. Hickey, M. van Andel, and Z. Yu, 2017: Evaluating the benefits of vaccination when used in combination with stamping-out measures against hypothetical introductions of foot-and-mouth disease into New Zealand: a simulation study. *N. Z. Vet. J.* **65**, 124–133, DOI: 10.1080/00480169.2016.1263165.

Seibel, R.L., A.J. Meadows, C. Mundt, and M. Tildesley, 2024: Modeling target-density-based cull strategies to contain foot-and-mouth disease outbreaks. *PeerJ* **12**, e16998, DOI: 10.7717/peerj.16998.

SENASAG, 2015: Manual de Procedimientos para la Atención de Sospechas de Enfermedades Vesiculares y Contingencia por Fiebre Aftosa [Online] Available at https://www.senasag.gob.bo/images/Contenido_Unidades/SA/Area_epidemiologia/Plan%20Contingencia%20fiebre%20aftosa.pdf (accessed October 9, 2024).

SENASAG, B. n.d.: Servicio Nacional de Sanidad Agropecuaria e Inocuidad Alimentaria, Inicio [Online] Available at https://www.senasag.gob.bo/ (accessed October 29, 2024).

Singh, R.K., G.K. Sharma, S. Mahajan, K. Dhama, S.H. Basagoudanavar, M. Hosamani, B.P. Sreenivasa, W. Chaicumpa, V.K. Gupta, and A. Sanyal, 2019: Foot-and-Mouth Disease Virus: Immunobiology, Advances in Vaccines and Vaccination Strategies Addressing Vaccine Failures—An Indian Perspective. *Vaccines* **7**, 90, DOI: 10.3390/vaccines7030090.

Stenfeldt, C., M. Eschbaumer, S.I. Rekant, J.M. Pacheco, G.R. Smoliga, E.J. Hartwig, L.L. Rodriguez, and J. Arzt, 2016: The Foot-and-Mouth Disease Carrier State Divergence in Cattle. (S. Perlman, Ed.)*J. Virol.* **90**, 6344–6364, DOI: 10.1128/JVI.00388-16.

Tildesley, M.J., and M.J. Keeling, 2008: Modelling foot-and-mouth disease: A comparison between the UK and Denmark. *Prev. Vet. Med.* **85**, 107–124, DOI: 10.1016/j.prevetmed.2008.01.008.

Tomassen, F.H.M., A. De Koeijer, M.C.M. Mourits, A. Dekker, A. Bouma, and R.B.M. Huirne,

2002: A decision-tree to optimise control measures during the early stage of a foot-and-mouth disease epidemic. *Prev. Vet. Med.* **54**, 301–324, DOI: 10.1016/S0167-5877(02)00053-3.

Traulsen, I., G. Rave, J. Teuffert, and J. Krieter, 2011: Consideration of different outbreak conditions in the evaluation of preventive culling and emergency vaccination to control foot and mouth disease epidemics. *Res. Vet. Sci.* **91**, 219–224, DOI: 10.1016/j.rvsc.2010.12.016.

Ulziibat, G., E. Raizman, A. Lkhagvasuren, C.J.M. Bartels, O. Oyun-Erdene, B. Khishgee, C. Browning, D.P. King, A.B. Ludi, and N.A. Lyons, 2023a: Comparison of vaccination schedules for foot-and-mouth disease among cattle and sheep in Mongolia. *Front. Vet. Sci.* **10**, 990043, DOI: 10.3389/fvets.2023.990043.

Ulziibat, G., E. Raizman, A. Lkhagvasuren, C.J.M. Bartels, O. Oyun-Erdene, B. Khishgee, C. Browning, D.P. King, A.B. Ludi, and N.A. Lyons, 2023b: Comparison of vaccination schedules for foot-and-mouth disease among cattle and sheep in Mongolia. *Front. Vet. Sci.* **10**, 990043, DOI: 10.3389/fvets.2023.990043.

Virtanen, P., R. Gommers, T.E. Oliphant, M. Haberland, T. Reddy, D. Cournapeau, E. Burovski, P. Peterson, W. Weckesser, J. Bright, S.J. van der Walt, M. Brett, J. Wilson, K.J. Millman, N. Mayorov, A.R.J. Nelson, E. Jones, R. Kern, E. Larson, C.J. Carey, İ. Polat, Y. Feng, E.W. Moore, J. VanderPlas, D. Laxalde, J. Perktold, R. Cimrman, I. Henriksen, E.A. Quintero, C.R. Harris, A.M. Archibald, A.H. Ribeiro, F. Pedregosa, P. van Mulbregt, and SciPy 1.0 Contributors, 2020: SciPy 1.0: Fundamental Algorithms for Scientific Computing in Python. *Nat. Methods* **17**, 261–272, DOI: 10.1038/s41592-019-0686-2.

Wickham, H., M. Averick, J. Bryan, W. Chang, L. McGowan, R. François, G. Grolemund, A. Hayes, L. Henry, J. Hester, M. Kuhn, T. Pedersen, E. Miller, S. Bache, K. Müller, J. Ooms, D. Robinson, D. Seidel, V. Spinu, K. Takahashi, D. Vaughan, C. Wilke, K. Woo, and H. Yutani, 2019: Welcome to the Tidyverse. *J. Open Source Softw.* **4**, 1686, DOI: 10.21105/joss.01686.

**Simulating foot-and-mouth dynamics and control in Bolivia**

## Model formulation and description

We developed a transmission model that integrates multiple hosts and a single pathogen at different scales to replicate the trajectories of FMD epidemics (Garira, 2018; Cespedes Cardenas and Machado, 2024; Cespedes Cardenas et al., 2024) and allows for simulating countermeasures. This model resulted in the creation of a software package named "MHASpread: A multi-host animal spread stochastic multilevel model" (version 3.0.0), which is available for more detailed information at https://github.com/machado-lab/MHASPREAD-model. MHASpread enables the explicit definition of transmission probabilities specific to each species and the periods when the disease can be transmitted to multiple species. At the level of individual farms, the model accounts for each species' birth and death data.

## Within-farm dynamics

For the within-farm dynamics, we assume populations were homogeneously distributed. Species were homogeneously mixed in farms with at least two species, meaning that the probability of contact among species was homogeneous regardless of/when species were segregated in barns and/or paddocks (e.g., commercial swine farms are housed in barns with limited changes of direct contact with cattle). The within-farm dynamics consist of mutually exclusive health states (i.e., an individual can only be in one state per discrete time step) for animals of each species (bovines, swine, and small ruminants). Health states (hereafter, “compartments”) include susceptible $(S)$, exposed $(E)$, infectious, $(I)$, and recovered $(R)$, defined as follows:

Susceptible: animals that are not infected and are susceptible to infection.

Exposed: animals that have been exposed but are not yet infected.

Infectious: infected animals that can successfully transmit the infection.

Recovered: animals that have recovered and are no longer susceptible.

Our model considers birth and death, which is used to update the population of each farm. The total population is calculated as $N = S + E + I + R$. The number of individuals within each compartment transitions from $S\beta \rightarrow\ E, \frac{1}{\sigma} \rightarrow\ I, \frac{1}{\gamma} \rightarrow\ R$ according to the following equations:

$$\frac{dS_i(t)}{dt} = u_i(t) - v_i(t) - \frac{\beta S_i(t) I_i(t)}{N_i} \quad (1)$$

$$\frac{dE_i(t)}{dt} = \frac{\beta S_i(t) I_i(t)}{N_i} - v_i(t)E_i(t) - \frac{1}{\sigma}E_i(t) \quad (2)$$

$$\frac{dI_i(t)}{dt} = \frac{1}{\sigma}E_i(t) - \frac{1}{\gamma}I_i(t) - v_i(t)I_i(t) \quad (3)$$

$$\frac{dR_i(t)}{dt} = \frac{1}{\gamma}I_i(t) - v_i(t) \quad (4)$$

Transmission depends on infected and susceptible host species, as reflected by the species-specific FMD transmission coefficient $\beta$ (Table S1).

**Supplementary Table S1**. The distribution of each host-to-host transmission coefficient (β) per animal$^{-1}$ day$^{-1}$.

| **Infected species** | **Susceptible taxon** | **Transmission coefficient (β), shape and distribution (min, mode, max)** | **Reference** |
|---|---|---|---|
| Bovine | Bovine | PERT (0.18, 0.24, 0.56) | Calculated from the 2000-2001 FMD outbreaks in the state of Rio |
| | | | Grande do Sul (da Costa et al., 2022) |
| Bovine | Swine | PERT (0.18, 0.24, 0.56) | Assumed |
| Bovine | Small ruminants | PERT (0.18, 0.24, 0.56) | Assumed |
| Swine | Bovine | PERT (3.7, 6.14, 10.06) | Assumed (Eblé et al., 2006) |
| Swine | Swine | PERT (3.7, 6.14, 10.06) | (Eblé et al., 2006) |
| Swine | Small ruminants | PERT (3.7, 6.14, 10.06) | Assumed (Eblé et al., 2006) |
| Small ruminants | Bovine | PERT (0.044, 0.105, 0.253) | Assumed (Orsel et al., 2007) |
| Small ruminants | Swine | PERT (0.006, 0.024, 0.09) | (Goris et al., 2009) |
| Small ruminants | Small ruminants | PERT (0.044, 0.105, 0.253) | (Orsel et al., 2007) |

Bolivian data do not consistently record the number of animal births and deaths, (excluding those sent to slaughterhouses). Despite this limitation, our model can incorporate these births and deaths whenever data are available. Here, births are represented by the number of animals born alive $u_i(t)$ that enter the $S$ compartment on the farm $i$ at the time $t$ according to the day-to-day records; similarly, $v_i(t)$ represent the exit of the animals from any compartment due to death at the time $t$. The transition from $E$ to $I$ is driven by $1/\sigma$, and the transition from $I$ to $R$ is driven by $1/\gamma$; these values are drawn from the distribution generated from each specific species according to the literature (Supplementary Material Table S2).

**Supplementary Table S2**. The within-farm distribution of latent and infectious FMD parameters for each species.

| **FMD parameter** | **Species** | **Mean, median (25th, 75th percentile) in days** | **Reference** |
|---|---|---|---|
| Latent period, $\sigma$ | Bovine | 3.6, 3 (2, 5) | (Mardones et al., 2010) |
| | Swine | 3.1, 2 (2, 4) | (Mardones et al., 2010) |
| | Small ruminants | 4.8, 5 (3, 6) | (Mardones et al., 2010) |
| Infectious period, $\gamma$ | Bovine | 4.4, 4 (3, 6) | (Mardones et al., 2010) |
| | Swine | 5.7, 5 (5, 6) | (Mardones et al., 2010) |

| | | |
|---|---|---|
| Small ruminants | 3.3, 3 (2, 4) | (Mardones et al., 2010) |

Note: The time unit is days.

**Kernel transmission dynamics**

Spatial transmission encompasses a range of mechanisms, such as airborne transmission, animal contact over fence lines, and equipment sharing between farms (Boender et al., 2010; Boender and Hagenaars, 2023). Local spread was modeled using a spatial transmission kernel, in which the likelihood of transmission decreased as a function of the between-farm distance. The probability $PE$ at time $t$ describes the likelihood that a farm becomes exposed and is calculated as follows:

$$PE_j(t) = 1 - \prod_i \left(1 - \frac{I_i(t)}{N_i} \varphi e^{-\alpha d_{ij}}\right) \tag{5}$$

where $j$ represents the uninfected population and $ad_{ij}$ represents the distance between farm $j$ and infected farm $i$, with a maximum of 40 km (Cespedes Cardenas et al., 2024). Given the extensive literature on distance-based FMD dissemination and a previous comprehensive mathematical simulation study (Björnham et al., 2020), distances above 40 km were not considered. Here, $1 - \frac{I_i(t)}{N_i} \varphi e^{-\alpha d_{ij}}$ represents the probability of transmission between farms $i$ and $j$ scaled by infection prevalence of farm $i$, $\frac{I_i}{N_i}$, given the distance between the farms in kilometers. The parameters $\varphi$ and α control the shape of the transmission kernel; $\varphi = 0.044$, which is the probability of transmission when $d_{ij} = 0$, and $\alpha = 0.6$ control the steepness with which the probability declines with distance (Boender et al., 2010; Boender and Hagenaars, 2023; Cespedes Cardenas et al., 2024).

## FMD spread and control actions

We initially simulated a silent spread over 20 days, generating a broad spectrum of outbreak scenarios before implementing any control measures, as shown in Figure 2. The FMD control scenarios include the following measures: i) depopulation of infected farms, ii) emergency vaccination of farms within the infected and buffer zones, iii) a 30-day standstill on animal movement, and iv) the establishment of three distinct control zones around infected farms: a 3 km infected zone, a 7 km buffer zone, and a 15 km surveillance zone (Supplementary Material Figure S3).

The depopulation of infected farms involves destroying all animals from farms within the infected zone(s), with priority given to farms that have larger animal populations. Once depopulated, these farms are excluded from the simulation. When the daily depopulation capacity cannot accommodate all infected farms in one day, the remaining farms are scheduled for depopulation the following day or as soon as possible, subject to each simulated scenario's capacity limits (Table 2).

*Vaccination:* Bovine farms within the infected and buffer zones received emergency vaccination 15 days after establishing the control zone. Due to capacity constraints, farms unable to be vaccinated within a day were vaccinated on subsequent days (s).

The number of farms vaccinated per day depends on the scenario being simulated (Table 2). Vaccination is administered to farms within the infected and/or buffer zones, following predefined criteria, with priority given to farms with the largest population. Each farm's vaccination begins on a specified day, with animals progressively transferred from the $S$, $E$, $I$, and $R$ compartment to the $V$ compartment at a daily rate proportional to the remaining population in each compartment. The transfer is based on the vaccination rate and vaccine efficacy, with a maximum vaccination protection of 90% that was achieved within 15 days

The total number of animals eligible for vaccination is given by:

$$n = (1 - ve) * N \tag{6}$$

where $ve$ represents the vaccine efficacy and $N$ is the total number of cattle on the vaccinated farm. From the eligible to be vaccinated population, the number of animals moved to the vaccinated status each day is drawn from a binomial distribution:

$$X \sim Binomial(n, p)$$

where the probability $p$ is defined as:

$$p = \frac{1}{vt}$$

Here, $vt$ represents the time required for the full progression of immunity following vaccination (15 days), assuming that by the end of this period, all eligible animals will have developed immunity.

*Traceability*: We employed contact tracing to identify farms that had direct contact with infected farms in the past 30 days. These farms were subject to surveillance, including clinical examinations and detection procedures. Farms that tested positive during contact tracing were classified as detected infected farms.

*Infected farms detection:* We assumed that 10% of the infected farms were detected at the start of control measures, which occurred 10 days after the introduction of the index case. For example, if 100 farms were infected initially, 10 farms were detected. If the number of detected farms was less than one, we rounded up to one detected farm.

The detection rate was influenced by two main factors: the total number of farms within the control zones and the number of infected farms. For instance, when fewer farms were under surveillance but there was a higher number of infected farms, the likelihood of detection increased (Supplementary Material Figure S3). Infected farms outside the control zones were also included among those subjected to detection.

The probability of detecting a new infected farm $P_i$ considers the number of infected farms and the total number of farms under surveillance.

-$P_i$ : Number of farms in surveillance.

- $I_i$: Number of infected farms.

- $E$: Number of farms found in the current iteration

The algorithm operates as follows:

1. If $I_i < 5$ , $E$ is set to a random value between 0 and $I_i$ (inclusive), i.e.,$E = sample(0{:}\, I_i, 1)$.

2. If $Ii \geq 5$ a probability distribution is calculated using the hypergeometric distribution with parameters $(m = I_i)$, $(n = P_i)$ and $(k = \frac{p_i}{3})$.

Sample a value $p$ from this probability distribution.

Calculate $E = [I_i * p]$

If $E = 0$, set $E = 1$

*Traceability and movement standstill:* We employed contact tracing to detect farms with direct links to infected farms within the past 30 days, subjecting these farms to surveillance. Positive farms from trace back were categorized as detected infected farms, triggering the application of the same criteria for control zones. In addition, a 30-day restriction on animal movement was enforced across all three control zones, prohibiting both incoming and outgoing movements. The control zones remained in place, and the movement standstill was maintained until depopulation efforts were fully completed.

***References***

Björnham, O., R. Sigg, and J. Burman, 2020: Multilevel model for airborne transmission of foot-and-mouth disease applied to Swedish livestock. (Bryan C. Daniels, Ed.)*PLOS ONE* **15**, e0232489, DOI: 10.1371/journal.pone.0232489.

Boender, G.J., and T.J. Hagenaars, 2023: Common features in spatial livestock disease transmission parameters. *Sci. Rep.* **13**, 3550, DOI: 10.1038/s41598-023-30230-w.

Boender, G.J., H.J.W. van Roermund, M.C.M. de Jong, and T.J. Hagenaars, 2010: Transmission risks and control of foot-and-mouth disease in The Netherlands: Spatial patterns. *Epidemics* **2**, 36–47, DOI: 10.1016/j.epidem.2010.03.001.

Cespedes Cardenas, N., F. Amadori Machado, C. Trois, V. Maran, A. Machado, and G. Machado, 2024: Modeling foot-and-mouth disease dissemination in Rio Grande do Sul, Brazil and evaluating the effectiveness of control measures. *Front. Vet. Sci.* **11**.

Cespedes Cardenas, N., and G. Machado, 2024: MHASPREAD model [Online] Available at https://github.com/machado-lab/MHASPREAD-model (accessed September 16, 2024).

da Costa, J.M.N., L.G. Cobellini, N.C. Cardenas, F.H.S. Groff, and G. Machado, 2022:

Assessing epidemiological parameters and dissemination characteristics of the 2000 and 2001 foot-and-mouth disease outbreaks in Rio Grande do Sul, Brazil. *bioRxiv*2022.05.22.492961, DOI: 10.1101/2022.05.22.492961.

Eblé, P., A. De Koeijer, A. Bouma, A. Stegeman, and A. Dekker, 2006: Quantification of within- and between-pen transmission of Foot-and-Mouth disease virus in pigs. *Vet. Res.* **37**, 647–654, DOI: 10.1051/vetres:2006026.

Garira, W., 2018: A primer on multiscale modelling of infectious disease systems. *Infect. Dis. Model.* **3**, 176–191, DOI: 10.1016/j.idm.2018.09.005.

Goris, N.E., P.L. Eblé, M.C.M. de Jong, and K.D. Clercq, 2009: Quantifying foot-and-mouth disease virus transmission rates using published data. *ALTEX - Altern. Anim. Exp.* **26**, 52–54, DOI: 10.14573/altex.2009.1.52.

Mardones, F., A. Perez, J. Sanchez, M. Alkhamis, and T. Carpenter, 2010: Parameterization of the duration of infection stages of serotype O foot-and-mouth disease virus: an analytical review and meta-analysis with application to simulation models. *Vet. Res.* **41**, 45, DOI: 10.1051/vetres/2010017.

Orsel, K., A. Dekker, A. Bouma, J.A. Stegeman, and M.C.M. De Jong, 2007: Quantification of foot and mouth disease virus excretion and transmission within groups of lambs with and without vaccination. *Vaccine* **25**, 2673–2679, DOI: 10.1016/j.vaccine.2006.11.048.

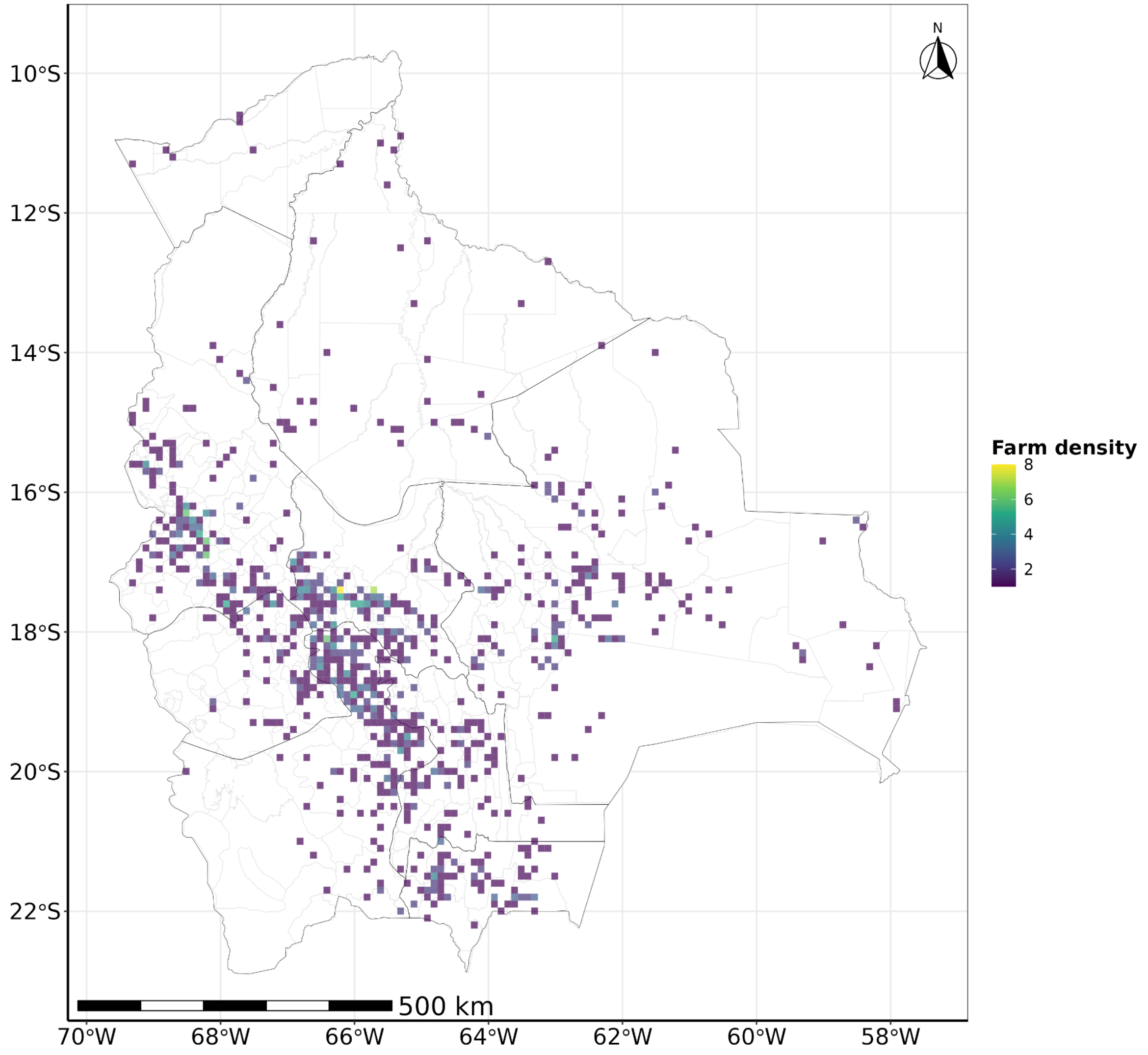

**Supplementary Figure S1.** The grid density of the sampled farms to seed initial outbreaks. A 10 $km^2$ grid was projected onto the map, and the density of pig premises was represented as the number of premises allocated in each grid cell. Black lines represent the political state division of Bolivia. The gray lines represent the political municipal division of Bolivia.

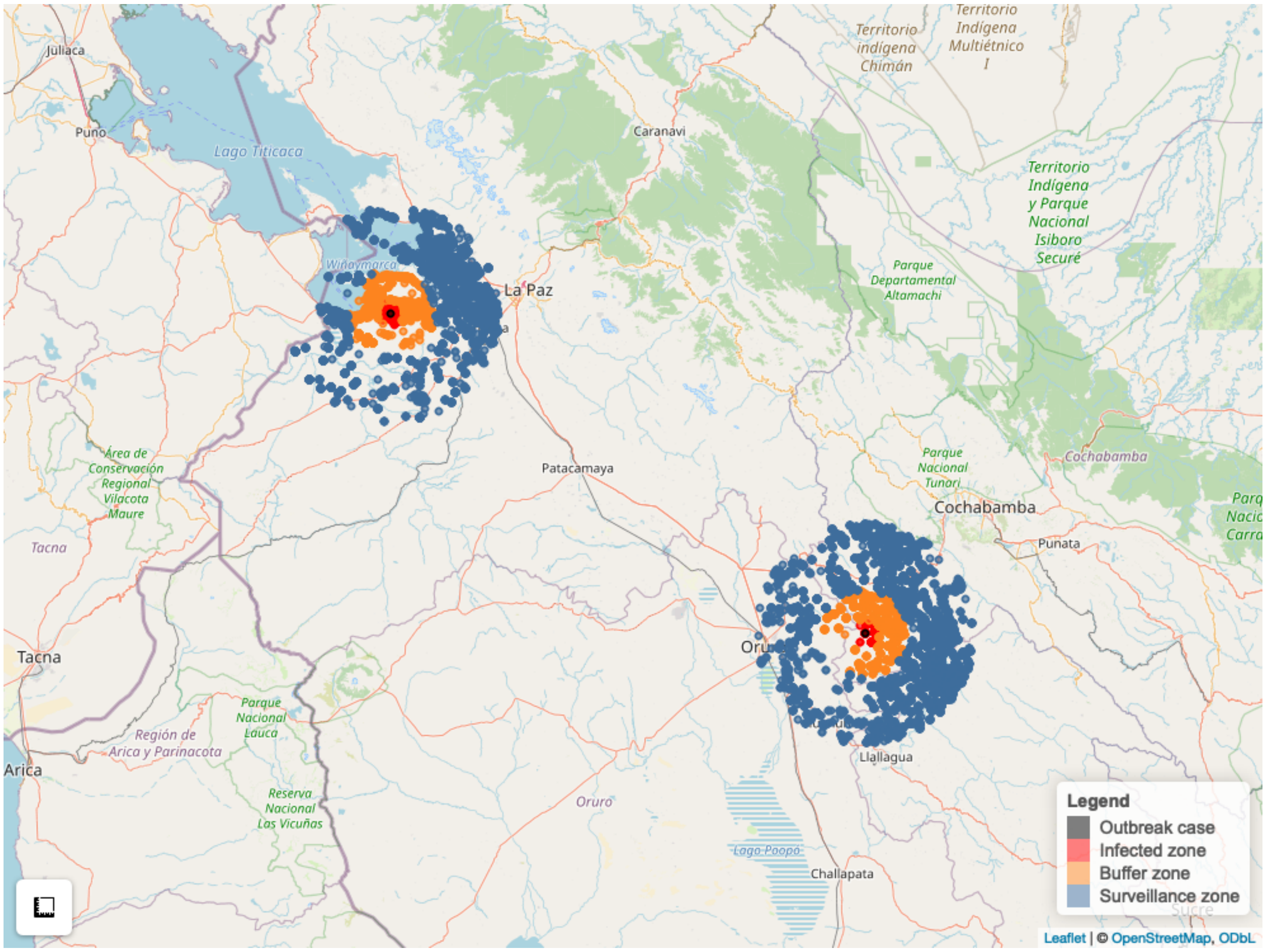

**Supplementary Figure S2. Control zones mapping.** The infected control zones of 3 km are represented as red dots, the 7 km buffer zone is represented as orange dots, and the 15 km surveillance zone is represented as blue dots.

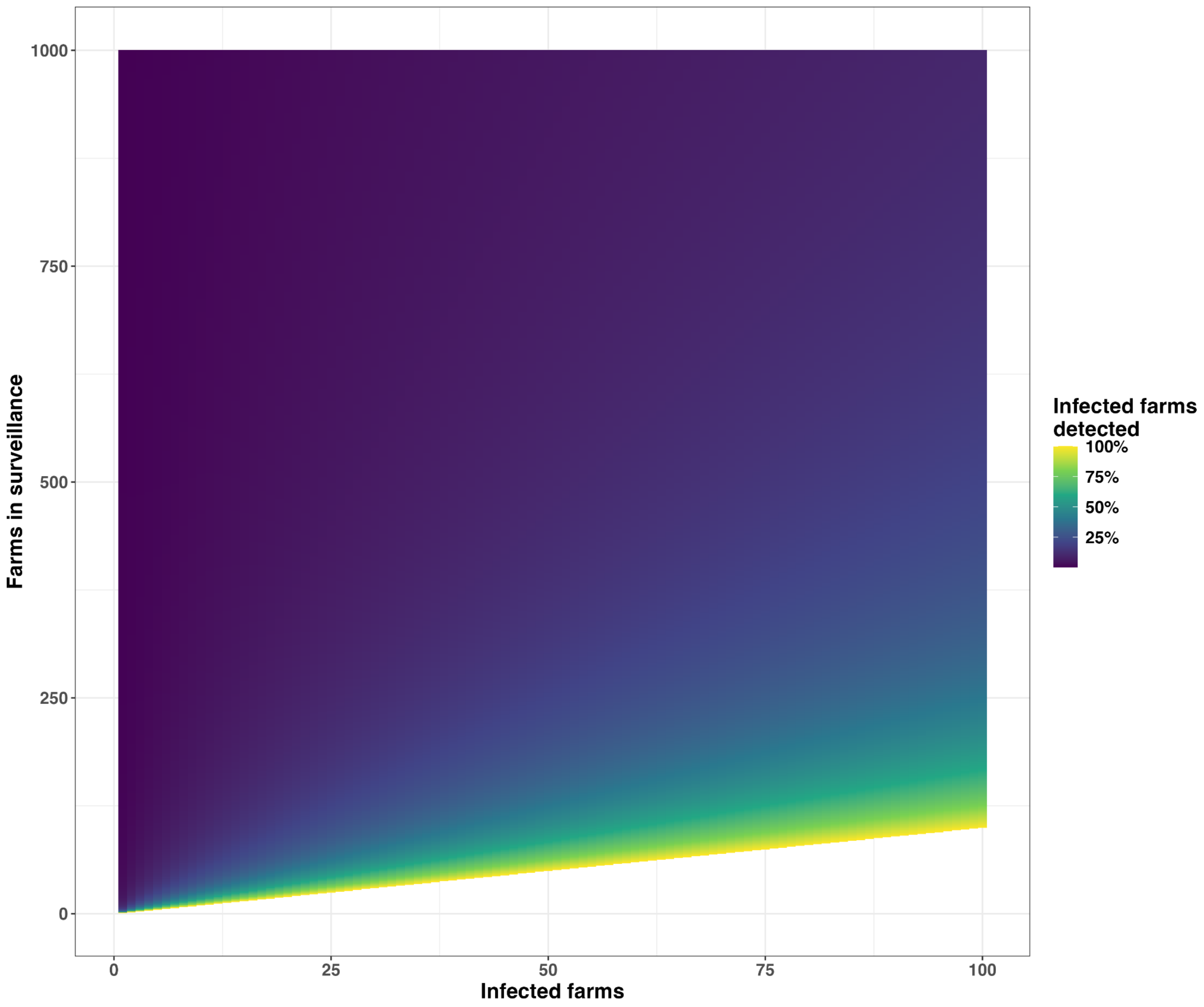

**Supplementary Figure S3. Percentage of detected farms according to prevalence and population size**. The y-axis represents the number of farms under surveillance, while the x-axis represents the number of infected farms in the population. The color represents the percentage of infected farms that will be detected. Figure from (Cespedes Cardenas et al., 2024).